\documentstyle[12pt]{article}
\begin{document}
\vskip 2 cm
\begin{center}
\Large{\bf ON THE NUMBER OF COLOURS IN QUANTUM CHROMODYNAMICS }
\end{center}
\vskip 3 cm
\begin{center}
{\bf A. Abbas} \\
Department of Physics and Astronomy\\
University of Pittsburgh, Pittsburgh PA 15260\\
(e-mail: abbas@phyast.pitt.edu)\\
and\\
Institute of Physics, Bhubaneswar-751005, India \\
(e-mail : afsar@iopb.res.in)
\end{center}
\vskip 20 mm  
\begin{centerline}
{\bf Abstract }
\end{centerline}
\vskip 3 mm

It is commonly believed that  ${\pi}^0 \rightarrow 2 \gamma$ 
decay shows that there are three colours in Quantum Chromodynamics
(QCD). It is shown here that this is not correct.  When correct 
colour dependent charges of the quarks are considered then
it is shown that this decay does not make any statement about 
the number of colours in QCD.

\newpage

It is important to know how many colours there are so that one may
confidently formulate the theory of the strong interaction, ie Quantum
Chromodynamivcs ( QCD ) around it. Fortunately there are several
experimental indications that the number of colours in nature are three.
In the authors view the best indication comes from the experimental
evaluation of the R - ratio in the reaction

\begin{equation}
e^+ + e^- \rightarrow hadrons
\end{equation}

However it is very often stated that  
even a more important proof of the number of colours being three
comes from the experimental analysis of the reaction
$\pi^o \rightarrow 2 \gamma$ . Although this occurs in several 
places,  for the sake of example here we would like to bring out this 
fact from references [1 - 4].
The aim of this short paper is to prove that this is not correct.

The decay  $\pi^o \rightarrow 2 \gamma$
takes place through the anomaly in the divergence of axial vector
currents [ 1 - 4 ]. The decay rate [ 1 - 4 ] is

\begin{equation}
\Gamma(\pi^0 \rightarrow 2 \gamma)={N_c}^2 {({Q_u}^2-{Q_d}^2)}^2
 \frac{\alpha^2 {m_{\pi^0}}^3}{64 \pi^3 {F_{\pi}}^2}
\end{equation}

where $Q_u$ and $Q_d$ are the u- and d- quark charges ( in units of the
proton charge ), $N_c$ is the number of colours, 
$ m_{\pi_0} $ is the neutral pion
mass, $\alpha$ = $\frac {e^2} {4 \pi} $ and $F_{\pi}$, the pion decay
constant is taken as 91 MeV.

The experimental value of the decay rate is around 7.48 eV.
The quark charges are canonically taken as $Q_u$=2/3 and $Q_d$=-1/3
[ 1 - 4 ]. If there were no colours , ie $N_c$ = 1 then from the above
formula one gets the decay rate to be 0.84 eV. This is way off. Only when
one takes the number of colours to be 3 in the above formula 
it is then that one obtains a satisfactory agreement with the experiment.
This is then taken as a clean proof of three colours in QCD [ 1 - 4 ].

Now the factor 

\begin{equation}
{N_c}^2 {( {Q_u}^2 - {Q_d}^2 )}^2
\end{equation}

is the one which presumably determines the number of colours in the pion
decay rate above. As shown above, for the u- and the d- quark charges 2/3
and -1/3 respectively, it is 1/9 when quarks are uncoloured and equal to
one ( and giving agreement with the experiment ) when $N_c$=3.

However, in the above one  has taken the u- and the d- quark 
electric charges to be static at 2/3 and -1/3 respectively. 
By the word 'static' we  mean that they remain the same whether the number
of colours were 1 or 3 or 5 or any other number. What we'd like to point
out is that this is not correct. And this is the source of the problem
here. As has been shown  by the author, the electric charges of quarks
have a colour dependence  and the correct charges for the quarks
are reproduced for three colours [ 5,6 ].

It has been shown by the author [ 5,6 ] that in the Standard Model when
the  $ SU(N_{c}) \otimes SU(2)_{L} \times U(1)_{Y} $
( for $ N_c = 3 $ )  symmetry is 
spontaneous breaken by a Higgs mechanism to
$ SU(N_c) \otimes U(1)_{em} $ symmetry
and by maintaining all the built-in properties of the standard Model one
finds the correct electric charges of the quarks as [ 5,6 ] :

\begin{equation}
Q_u = \frac{1}{2} ( 1 + \frac{1}{N_c} )
\end{equation}

\begin{equation}
Q_d = \frac{1}{2} ( -1 + \frac{1}{N_c} )
\end{equation}

For $N_c$ = 3 these reproduce the correct charges for the u- and 
the d- quarks as determined experimentally.
That these are actually the correct charges and that they represent the
correct colour dependence  of the electric charge has been amply
demonstarted by the author [ 5-9 ]. In fact this should be treated
as a unique property and prediction of the Standard Model.

Note that as shown earlier it is the factor
$ {N_c}^2 {( {Q_u}^2 - {Q_d}^2 ) }^2 $ in expression 3 above
in the decay rate of the pion which determined the number of colours to
be three. However, with the correct colour dependent charges found by the
author ( equations  4 and 5 above ) this factor is found to be always 
one ( and thus giving good fit to the experiment ),
irrespective of the number of colours.
This means that the decay rate of the pion gives good fit to the
experiment for any number of colours. Thus clearly the decay rate of pion
does not determine the number of colours. 
Hence, these are best determined by the other
experimental methods as is well documented  [1-4] but certainly 
not through the pion decay analysis ( as shown here ).

In short, we have shown that in contrast to what is  
commonly believed and   which is well documented [ 1 - 4 ], 
the electric charges of the u- and the d- quarks are not just the static
2/3 and -1/3 respectively for any number of colours. 
The electric charges of the quarks
have colour dependence [ 5 - 6 ] and this is what has been used here.
When these correct colour dependent charges are used
for the decay rate of ${\pi}^0 \rightarrow {2 \gamma}$
then one finds that it has nothing to say about the number of colours
in QCD. Hence the popular misconception as to this decay having anything
to say about colour has to be rectified. The basic mistake that was made
was not to realize that the electric charge,
as arising in the Standard Model, has a colour dependence. This fact is
important as QCD in the limit of $ N_c $ going to infinity
continues to play a significant role for a proper understanding
of the theory of the strong interaction.
 
\newpage

\begin{center}
{\bf\large REFERENCES }
\end{center}

\vskip 1 cm

1. T. Muta,
" Foundations of Quantum Chromodynamics ", World Scientific,
Singapore, 1987

\vskip 1 cm

2. Ta-Pei Cheng and Ling-Fong Li,
" Gauge Theory of Elementary Particle Physics ", Clarendon Press,
Oxford, 1989

\vskip 1 cm

3. F. E. Close,
" An Introduction to Quarks and Partons ", Academic Press, London, 1979

\vskip 1 cm

4. J. E. Dodd,
" The Ideas of Particle Physics ", Cambridge University Press,
Cambridge, 1984

\vskip 1 cm
 
5. A. Abbas, 
{\it Phys. Lett. } {\bf B 238} (1990) 344

\vskip 1 cm
 
6. A. Abbas,
{\it J. Phys. } {\bf G 16 } (1990) L163

\vskip 1 cm

7. A. Abbas,
{\it Nuovo Cim.}  {\bf A 106}, (1993) 985

\vskip 1 cm

8. A. Abbas,
{\it Physics Today }, July 1999, p.81-82

\vskip 1 cm

9. A. Abbas, {\it `Phase transition in the early universe and charge
quantization'}; hep-ph/9503496

\end{document}